\documentclass[10pt,conference]{IEEEtran}
\IEEEoverridecommandlockouts

\usepackage{cite}
\usepackage{amsmath,amssymb,amsfonts}
\usepackage{graphicx}
\usepackage{textcomp}
\usepackage{xcolor}
\usepackage{algorithm}
\usepackage{algpseudocode}
\def\BibTeX{{\rm B\kern-.05em{\sc i\kern-.025em b}\kern-.08em
    T\kern-.1667em\lower.7ex\hbox{E}\kern-.125emX}}
\begin{document}

\title{CrossTrace: Efficient Cross-Thread and Cross-Service Span Correlation in Distributed Tracing for Microservices}

\author{\IEEEauthorblockN{Linh-An Phan\IEEEauthorrefmark{1},
MingXue Wang\IEEEauthorrefmark{1},
Guangyu Wu \IEEEauthorrefmark{1}, 
Wang Dawei \IEEEauthorrefmark{2},
Chen Liqun \IEEEauthorrefmark{2}, and
Li Jin \IEEEauthorrefmark{2}}
\IEEEauthorblockA{\IEEEauthorrefmark{1}Huawei Ireland Research Center, Dublin, Ireland}
\IEEEauthorblockA{\IEEEauthorrefmark{2}Huawei Shenzhen R\&D Center, China}%
}

\maketitle

\begin{abstract}
Distributed tracing has become an essential technique for debugging and troubleshooting modern microservice-based applications, enabling software engineers to detect performance bottlenecks, identify failures, and gain insights into system behavior. However, implementing distributed tracing in large-scale applications remains challenging due to the need for extensive instrumentation. To reduce this burden, zero-code instrumentation solutions---such as those based on eBPF---have emerged, allowing span data to be collected without modifying application code. Despite this promise, span correlation---the process of establishing causal relationships between spans---remains a critical challenge in zero-code approaches. Existing solutions often rely on thread affinity, compromise system security by requiring the kernel integrity mode to be disabled, or incur significant computational overhead due to complex inference algorithms. This paper presents \textbf{\textit{CrossTrace}}, a practical and efficient distributed tracing solution designed to support the debugging of microservice applications without requiring source code modifications. CrossTrace employs a greedy algorithm to infer intra-service span relationships from delay patterns, eliminating reliance on thread identifiers. For inter-service correlation, CrossTrace embeds span identifiers into TCP packet headers via eBPF, enabling secure and efficient correlation compromising system security policies. Evaluation results show that CrossTrace can correlate thousands of spans within seconds with over 90\% accuracy, making it suitable for production deployment and valuable for microservice observability and diagnosis.
\end{abstract}

\begin{IEEEkeywords}
Distributed Tracing, eBPF, Zero-code Instrumentation, Observability, Microservices
\end{IEEEkeywords}

\section{Introduction}
\label{sec:1} 
Distributed tracing is a key pillar of the observability stack for modern microservices, offering insights into request paths and performance bottlenecks, and enabling faster root cause analysis of service failures. As modern applications increasingly adopt microservice architecture to improve modularity and development agility, the complexity of troubleshooting such systems has grown exponentially. Unlike monolithic applications, where all function calls are intra-process communications and logs or metrics may suffice for root cause analysis, microservices communicate via remote procedure calls. Therefore, having visibility into how requests propagate across components helps developers understand service topology, detect bottlenecks, and debug failures across services \cite{9678534,mystery,trace-abnormally,tracing-debug,tracing-debug2}. Distributed tracing provides this capability by tracking end-to-end request flows throughout the system.

However, implementing distributed tracing is time-consuming and error-prone, particularly for microservice applications. This process typically involves modifying application source code to insert tracing logic (e.g., instrumentation and trace context propagation) and subsequently redeploying the microservices. In large systems comprising hundreds of microservices written in different programming languages and maintained by separate teams, manual instrumentation imposes a major engineering burden and increases the risk of inconsistencies or omissions. As a result, non-intrusive approaches, commonly referred to as \textbf{zero-code instrumentation}, have emerged as attractive solutions to enable distributed tracing without manual changes to application code.

Two recent techniques for achieving zero-code instrumentation include: (1) using auto-instrumentation libraries provided by OpenTelemetry \cite{otel}, and (2) leveraging extended Berkeley Packet Filter (eBPF) technology. In the first technique, commonly used libraries (e.g., Spring Boot for Java or Django for Python) are instrumented in advance, allowing microservices that use these libraries to generate traces without additional application-level instrumentation. However, this approach depends on the availability of pre-instrumented libraries and requires updates to the runtime environment as well as redeployment of microservices. In contrast, eBPF enables the collection of detailed telemetry data from applications running on Linux systems by attaching monitoring programs to kernel functions (i.e., system calls or the network stack) \cite{ebpf-http}. Originally designed for packet filtering \cite{bpf}, eBPF has evolved into a versatile technology used in a wide range of scenarios, including networking, observability, and security \cite{ebpf-insight, ebpf-rise}.

\begin{figure}[!t]
	\centering
	\includegraphics[width=1.0\linewidth]{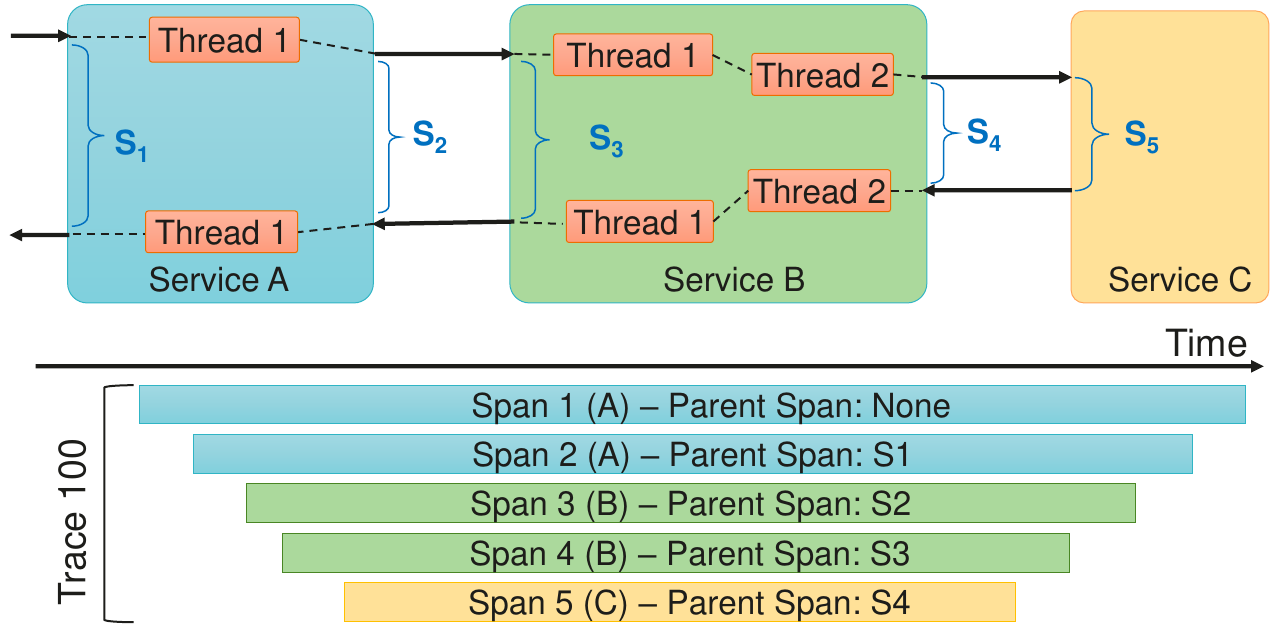}
	\caption{Complete trace (i.e., Trace 100) includes 5 spans (\(S_1\) to \(S_5\)) that represent the life-cycle of an end-to-end request traversing through three services. Parent span information indicates the causal relationship between two spans.}
	\label{fig1}
\end{figure}

In observability, eBPF is primarily used as a non-intrusive metrics collector due to its deep integration with the Linux kernel and low overhead. However, its application to distributed request tracing remains limited, with only a few open-source solutions, such as DeepFlow \cite{deepflow} and Grafana Beyla \cite{grafanabeyla}, exploring this capability. A primary challenge is correlating spans across distributed components to reconstruct complete request traces. This task is both crucial and non-trivial for eBPF-based tracing solutions, as it determines the accuracy and effectiveness of the resulting trace data. A span represents the duration of a single operation within an end-to-end tracing workflow. For example, spans \(S_3\) and \(S_5\) in Fig. \ref{fig1} represent the time between receiving and responding to a request in \textit{Service B} and \textit{Service C}, respectively. Capturing spans is merely the first step; correlating them to form a complete and correct trace is the key challenge. Span correlation refers to identifying causal (i.e., parent-child) relationships between spans and includes two types:
\begin{enumerate}
  \item \textbf{Intra-service (Intra-component) Correlation}: For example, linking span \(S_2\) to span \(S_1\) within \textit{Service A} and span \(S_4\) to span \(S_3\) within \textit{Service B} in Fig. \ref{fig1}.
  \item \textbf{Inter-service (Inter-component) Correlation}: For example, linking spans \(S_3\) of \textit{Service B} to \(S_2\) of \textit{Service A} and \(S_5\) of \textit{Service C} to \(S_4\) of \textit{Service B} in Fig. \ref{fig1}. 
\end{enumerate}

Trace context propagation (i.e., trace and span identifiers) is a common approach for \textbf{inter-service span correlation}. To implement this, Grafana Beyla modifies HTTP request headers and injects trace metadata using eBPF helper functions. However, this operation is restricted when the Linux kernel operates in integrity lockdown mode, which prevents eBPF programs from modifying kernel or user-space memory. As this mode is enabled by default in most Linux distributions, Beyla’s approach requires disabling kernel integrity, a practice that is strictly prohibited in most production environments due to its associated security risks \cite{beyla-limitation, ebpf-threat}. For \textbf{intra-service correlation}, both DeepFlow and Grafana Beyla assume that two spans executed by the same operating system (OS) thread belong to the same trace. While this thread-based assumption may hold for some services, it is frequently invalid in modern programming paradigms that rely on virtual threads (e.g., Goroutines in Golang \cite{goroutine} or Java virtual threads \cite{virtualthread}) or multi-threaded execution. In our production environment, modifying request headers---and thereby compromising system security---is entirely unacceptable, making these thread-based and header-modifying approaches fundamentally limited.

To address the limitations of DeepFlow and Grafana Beyla in \textbf{intra-service correlation}, a recent solution, TraceWeaver \cite{traceweaver}, attempts to infer span relationships based on delays between spans. It first constructs statistical distributions of delays between ingress and egress events within a service and then scores each potential correlation using probability density functions. An optimization algorithm is subsequently applied to select the most probable correlation. While TraceWeaver avoids thread-based assumptions, its runtime grows significantly with increasing request concurrency due to the computational cost of the optimization process.

In summary, current approaches to both intra-service and inter-service span correlation have critical limitations, making existing eBPF-based distributed tracing solutions impractical for deployment in production microservice systems. In this paper, we present \textbf{\textit{CrossTrace}}, an eBPF-based distributed tracing system designed to address the above limitations. The key innovations of CrossTrace are:
\begin{itemize}
\item \textbf{Greedy-Based Cross-Thread Span Correlation}: We propose a greedy algorithm for correlating spans within the same microservice, referred to as \textit{cross-thread span correlation}, as it does not rely on OS thread identifiers. Inspired by TraceWeaver, the algorithm leverages delay information to infer causality between ingress and egress spans. However, CrossTrace introduces several key enhancements: it employs adaptive thresholds to reduce candidate selection time, leverages easily-correlated requests to improve accuracy, and applies a greedy matching algorithm with conflict resolution to efficiently determine the most likely span correlations. These techniques enable fast and accurate correlation, making CrossTrace well-suited for high-load systems.

\item \textbf{eBPF-based Cross-Service Span Correlation}: We introduce a novel approach to propagate span IDs between two microservices by embedding them into TCP packet headers using eBPF helper functions with low overhead. This approach enables straightforward cross-service span correlation without modifying application-layer headers, thus preserving security settings, which is critical for production environments.
\end{itemize}

We conduct extensive evaluations of the CrossTrace prototype under varying load conditions, scaling up to over 1500 concurrent requests per microservice instance (i.e., container), surpassing the typical resource limitations of container services provided by leading cloud providers. The evaluation results show that CrossTrace achieves over 90\% correlation accuracy under high-load conditions and significantly outperforms a recent approach in terms of correlation runtime.

The rest of this paper is organized as follows. Section~\ref{sec:2} presents background knowledge and related work. Section~\ref{sec:3} describes the design of CrossTrace, and Section~\ref{sec:4} presents the evaluation results. While delay-only correlation may not achieve the same level of accuracy as instrumentation-based approaches, Section~\ref{sec:5} discusses how this limitation can be mitigated and how CrossTrace can still enhance the troubleshooting capabilities of microservice systems.


\begin{table*}[!tb]
    \centering
    \renewcommand{\arraystretch}{1.4}%
    \begin{tabular}{|p{1.8cm}|p{3.3cm}|p{4.8cm}|p{6.6cm}|}
        \hline
        \textbf{Solution} & \textbf{Span Generation} & \textbf{Span Correlation} & \textbf{Main Advantage/Limitation} \\ \hline
        OpenTelemetry \cite{otel} (auto instrumentation libraries) & Use instrumentation code in popular libraries or framework (e.g., Spring Boot, Django) to generate spans & Intra-service: Propagate trace context among function calls \newline Inter-service: Propagate trace context in request header & Advantage: High accuracy, generate traces in real-time \newline Limitation: Depend the availability of instrumented libraries, must re-deploy application after updating libraries \\ \hline
        Sidecar Proxy (e.g., Istio \cite{istio}, Linkerd \cite{Linkerd}) & Sidecar containers intercept all traffic and generate spans for request/response & Intra-service: Not supported \newline Inter-service: Propagate trace context in request header & Advantage: Support any types of application \newline Limitation: Cannot perform inter-service correlation. Add extra delays and resource overhead \\ \hline
        Deepflow \cite{deepflow} & Hook eBPF programs to kernel functions to capture and generate spans for requests/responses & Intra-service: Use OS thread identifier \newline Inter-service: Collect and compare network flow information to infer correlation & Advantage: Support old kernel versions since network flow information can be easily collected \newline Limitation: Only support applications use single-thread mode. Need a centralized service to collect and process data for inter-service correlation \\ \hline
        Grafana Beyla \cite{grafanabeyla} & Hook eBPF programs to kernel functions to capture and generate spans for requests/responses & Intra-service: Use OS thread identifier \newline Inter-service: Modify HTTP header to propagate trace context & Advantage: Generate traces in real-time if applications use single-thread mode \newline Limitation: Must disable kernel integrity feature, can lead to security risks \\ \hline
        TraceWeaver \cite{traceweaver} & Can use sidecar proxies or eBPF programs & Intra-service: Employ statistics analysis and an optimization algorithm to infer span relationship based on delays \newline Inter-service: Not supported & Advantage: Do not rely on thread ID assumption \newline  Limitation: Slow correlation time and accuracy depends on the distinctness of delays among spans. It is not clear how to perform inter-service correlation \\ \hline
        \textbf{CrossTrace} & Hook eBPF programs to kernel functions to capture and generate spans for requests/responses & Intra-service: Employ a greedy algorithm to infer span relationship based on delays \newline Inter-service: Propagate span ID in TCP header using eBPF helper functions & Advantage: Do not rely on thread ID assumption, comply with security policies, and fast correlation time \newline  Limitation:  Accuracy depends on the distinctness of delays among spans \\ \hline
    \end{tabular}
    \caption{Comparison of distributed tracing solutions that support zero-code instrumentation.}
    \label{table1}
\end{table*}

\section{Background and Motivation}
\label{sec:2}
A comprehensive distributed tracing system typically includes multiple layers: trace generation, collection, storage, and visualization \cite{sambasivan2014so}. This paper focuses on \textbf{trace generation}, which is particularly critical as it directly affects the system’s usefulness for debugging and observability. In essence, generating a complete trace involves two primary tasks: (1) \textbf{generating spans} (i.e., capturing relevant events in the tracing workflow) and (2) \textbf{correlating spans} (i.e., establishing the causal relationships between spans).

\subsection{Span Generation}
A span includes its start time, end time, and associated metadata for debugging, and it represents an individual operation within an end-to-end workflow. A straightforward way to generate spans is to instrument the application. Since the introduction of the industrial implementation of distributed tracing via Google Dapper \cite{dapper}, several open-source distributed tracing projects have emerged, including Zipkin \cite{zipkin}, Jaeger \cite{jaeger}, OpenTracing \cite{opentracing}, and OpenCensus \cite{opencensus}. In 2019, OpenCensus and OpenTracing merged to form OpenTelemetry \cite{otel}, an open-source framework aimed at reducing fragmentation across tracing tools and establishing a de facto standard for instrumentation. OpenTelemetry provides SDKs for manual instrumentation and pre-instrumented libraries for automatic instrumentation. However, instrumentation at scale remains a significant challenge for the widespread adoption of distributed tracing. Therefore, zero-code instrumentation techniques, which rely on external collectors (i.e., agents) to generate spans for requests and responses, have gained significant traction.

For instance, a sidecar container (proxy) can be deployed alongside each microservice instance to intercept all incoming and outgoing traffic. This sidecar proxy enables the collection of spans for each request or response passing through the microservice \cite{sidecartiwari2017}. Orchestration platforms like Kubernetes (K8s) \cite{Kubernetes} simplify the deployment and configuration of sidecar proxy for each microservice. However, there are notable drawbacks to the sidecar approach \cite{sidecar-drawback}. First, it does not support intra-service span correlation since it does not touch the application code. Second, it introduces additional latency into communication between microservices, as all traffic must pass through the sidecar proxies. Additionally, it consumes significant computational resources of cluster since a sidecar must be deployed for each instance (e.g., pod in K8s) of the microservices. Consequently, the number of sidecar proxies is equal to the number of pods deployed in the application.

Recently, eBPF technology has emerged as a lightweight and efficient alternative for collecting telemetry data, mitigating the resource consumption issues of sidecar containers. By hooking into system or application-level functions, eBPF programs can capture runtime activities (e.g., request send/receive events) from any process running on a Linux machine. In cloud environments, microservices are typically deployed as containers, which are treated as processes by the Linux OS. Therefore, an eBPF agent can generate spans and metrics (e.g., latency, throughput, resource usage) for microservices without intercepting network traffic, thereby eliminating communication latency overhead. Moreover, since only a single eBPF agent is needed per node and eBPF functions execute within the kernel, the overall hardware resource consumption is significantly lower than that of the sidecar approach \cite{ebpf-sidecar}.

\subsection{Span Correlation}
Two common strategies for span correlation in distributed tracing are \textit{metadata propagation} and \textit{black-box inference} \cite{sambasivan2014so}. Metadata propagation requires modifying system components to explicitly transmit trace metadata, and is predominantly used in instrumentation-based solutions \cite{xtrace, dapper, zipkin}. This metadata propagation approach enables accurate span correlation, but requires a standardized and interoperable trace format across system components. To address this issue, W3C has defined a standardized format for unifying trace data across different distributed tracing solutions \cite{w3c}. 

Sidecar-based solutions \cite{istio, Linkerd, envoy} and Grafana Beyla \cite{grafanabeyla} adopt metadata propagation for inter-service span correlation. Since sidecar proxies manage network communications, they can inject trace metadata into HTTP headers. Similarly, Grafana Beyla modifies requests at the kernel level to embed trace metadata. However, correlating only inter-service spans is insufficient to reconstruct a complete trace; thus, existing sidecar-based solutions still require modifications to the application code to support intra-service span correlation.

In contrast, black-box inference aims to establish span relationships based on logs or timing information without modifying the application code to propagate metadata. For example, Deepflow utilizes network flow data (e.g., TCP tuples, TCP sequences, and timing) for inter-service correlations. This approach can achieve highly accurate correlation due to the uniqueness of such information for service-to-service communications. For intra-service span correlation, Deepflow \cite{deepflow}, Grafana Beyla \cite{grafanabeyla}, and vPath \cite{vpath} rely on thread IDs that associate with ingress and egress requests. The accuracy of this method depends depends on whether the application consistently uses the same thread for processing requests.

Timing-based inference provides another method to correlate spans without relying on thread IDs \cite{wap5, timing-infer, traceweaver} (i.e., statistical inference). For example, TraceWeaver \cite{traceweaver} constructs statistical delay models between ingress and egress events, scoring candidate correlations based on probability density functions. An optimization algorithm is then employed to select the most likely correlation. While the accuracy of this technique depends on the distinctness of delays between spans, TraceWeaver has shown that it can provide useful observability results. However, its correlation time increases significantly under high load due to the computational cost of the optimization solver.

Table \ref{table1} summarizes and compares the designs and limitations of existing distributed tracing solutions that support zero-code instrumentation. Compared to these approaches, CrossTrace emphasizes practicality and production readiness by adhering to system security policies and avoiding thread-based assumptions. Notably, CrossTrace is designed to balance scalability and accuracy, making it particularly suitable for high-load environments. We believe that CrossTrace can streamline the debugging and performance tuning processes in microservice-based applications.

\begin{table*}[t!]
\caption{Information of ingress and egress requests stored in eBPF map.}
\centering
\label{table:tb2}
\renewcommand{\arraystretch}{1.20}%
\begin{tabular}{llccccccc}
\hline
 \textbf{No.} & \textbf{Socket Info (Remote, Local)} & \textbf{Syscall} & \textbf{Protocol} & \textbf{StreamID} & \textbf{PID} & \textbf{Timestamp} & \textbf{Span ID} & \textbf{Span Type}\\
\hline
1 & 20.1.1.1:5555, 10.0.0.1:8080 & recv & HTTP & N/A & 1 & 08:00:01.001 & S-01 & Ingress\\  
2 & 30.1.1.1:6666, 10.0.0.2:8080 & recv & HTTP & N/A & 2 & 08:00:01.001 & S-02 & Ingress\\  
3 & 15.1.1.2:80, 10.0.0.1:4444    & send & gRPC & 100 & 1 & 08:00:01.002 & S-03 & Egress \\ 
4 & 15.1.1.2:80, 10.0.0.1:4444    & send & gRPC & 102 & 1 & 08:00:01.003 & S-04 & Egress\\ 
5 & 15.1.1.2:80, 10.0.0.1:4444    & recv & gRPC & 100 & 1 & 08:00:01.007 & S-03  \\ 
6 & 20.1.1.1:5555, 10.0.0.1:8080 & send & HTTP & N/A & 1 & 08:00:01.008 & S-01 \\  
\hline
\end{tabular}
\end{table*}

\begin{figure}[!t]
	\centering
	\includegraphics[width=0.9\linewidth]{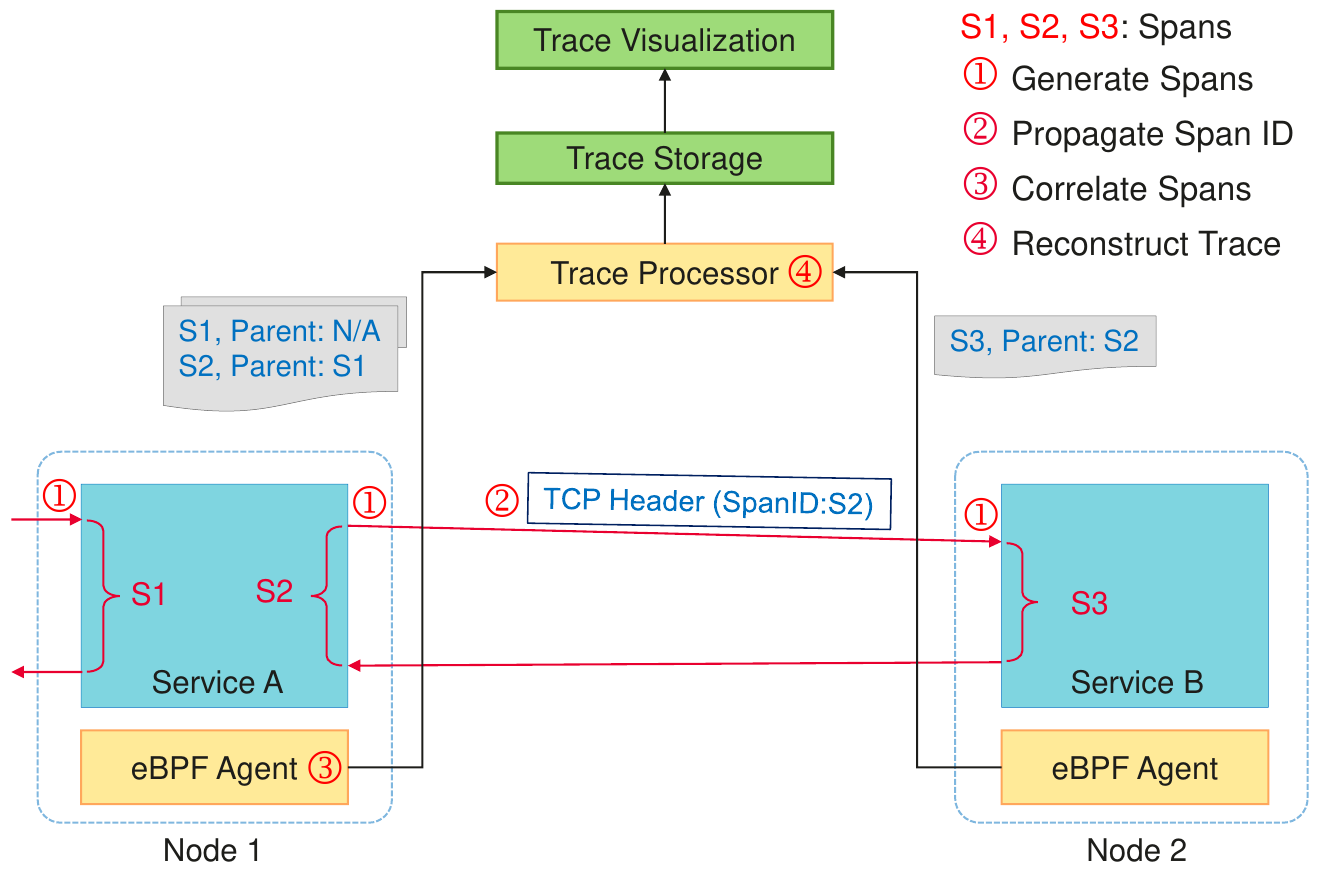}
	\caption{Overall architecture of Cross Trace, including two components as eBPF Agent and Trace Processor.}
	\label{fig2}
\end{figure}

\section{System Design}
\label{sec:3}
The design of CrossTrace comprises two main components: eBPF Agents and the Trace Processor service. eBPF Agents are deployed on each node in the cloud to generate and correlate local spans as shown in Fig. \ref{fig2}. To generate complete traces, four tasks are required as: (1) span generation, (2) cross-service span correlation, (3) cross-thread span correlation, and (4) trace reconstruction (i.e., assign a unique trace ID for correlated spans). The details of these tasks are described in the following subsections.



\subsection{Span Generation}
In distributed request tracing, a span represents the time taken to complete a request between two microservices. For a single request, two types of spans are captured:
\begin{itemize}
    \item \textbf{Ingress Span}: Represent the time from when a microservice receives a request to when it sends a response (e.g., spans S1 and S3 in Fig. \ref{fig2}).
    \item \textbf{Egress Span}: Represent the time from sending a request to receiving a response (e.g., span S2 in Fig. \ref{fig2}).
\end{itemize}
To generate these spans, it is necessary to capture the start time, end time, and relevant information of ingress and egress requests happening in the system. This is achieved by using eBPF programs attached to kernel system calls, such as \textit{send}, \textit{recv}, \textit{write}, and \textit{read}, which are invoked when a process sends or receives a request.

The eBPF programs collect and store metadata for each request and response, including socket information, request/response type, protocol, process identifier (PID), and timestamps, as summarized in Table \ref{table:tb2}. It is reasonable to assume that each microservice returns a response for every request, as spans can only be generated when both a request and its corresponding response are present. Based on the collected metadata, the eBPF programs perform lookups in eBPF maps to determine whether an event marks the start or end of a span. If the map contains an entry with matching socket information and protocol but with the opposite request/response type, the event is identified as the end of a span. Otherwise, the event marks the start of a new span. For example, rows 1 and 6 in Table \ref{table:tb2} represent the start and end times of an ingress span for Process ID 1, while rows 3 and 5 represent the start and end times of an egress span in the same process.

When an event is marked as the start of a span, the eBPF program generates a Span ID and stores the metadata in the eBPF map. If an event marks the end of a span, the span information is transferred from the kernel to the eBPF Agent for further processing, and the corresponding data are deleted from the eBPF map. A similar approach for collecting telemetry data and generating spans has been adopted in many eBPF-based observability solutions \cite{ebpf-http, ebpf-net, ebpf-ViperProbe}.

\begin{figure}[!b]
	\centering
	\includegraphics[width=0.7\linewidth]{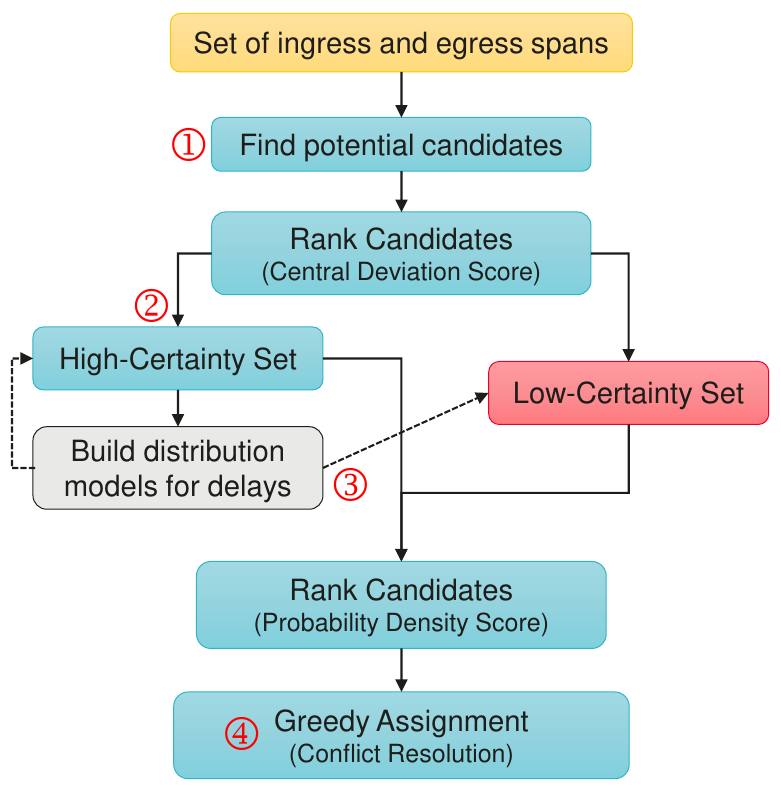}
	\caption{Steps of cross-thread span correlation process.}
	\label{fig3}
\end{figure}

\subsection{Cross-thread Span Correlation}
The fundamental idea behind \textit{Cross-thread Span Correlation} solution is to rely on delay pattern between spans. These insights are then used to evaluate potential candidates and determine the most appropriate correlations. The delays between spans can be classified into three types, which can be computed as follows:
\begin{itemize}
    \item The delay between the start time of the ingress span and the start time of the first egress span (e.g., \(d_1\) in Fig. \ref{fig4}).
    \item The delay between the end time of the ingress span and the end time of the last egress span (e.g., \(d_3\) and \(d_4\) in Fig. \ref{fig4}).
    \item The delay between two egress spans, which is the time from the end of the previous egress span to the start of the current one (e.g., \(d_2\) in Fig. \ref{fig4}).
\end{itemize}

To calculate these delays, the eBPF Agent requires the call graph of each microservice \cite{callgraph}, which represents the order and dependencies of spans, as illustrated in Fig. \ref{fig4}. It is important to note that the call graph information is attainable through different methods. It can either be constructed manually or automatically, for instance, by analyzing the source code or observing non-overlapping requests. In the case of non-overlapping requests, the ingress and egress spans clearly belong to the same request, making it straightforward to determine the call sequences and build the call graph.

Given a set of ingress and egress spans and a call graph of a microservice, the correlation process is performed through four steps, as shown in Fig.~\ref{fig3}:
\begin{enumerate}
\item Identify potential candidates for each ingress span.
\item Extract ingress spans with high confidence for determining correct correlations, referred to as the high-certainty set.
\item Build delay distribution models using the high-certainty set and then score all candidates based on a probability density function (PDF).
\item Perform greedy-based assignment to select the corresponding egress spans for each ingress span using PDF scores with conflict resolution.
\end{enumerate}
The details of each step are explained below:

\begin{figure}[!t]
	\centering
	\includegraphics[width=0.9\linewidth]{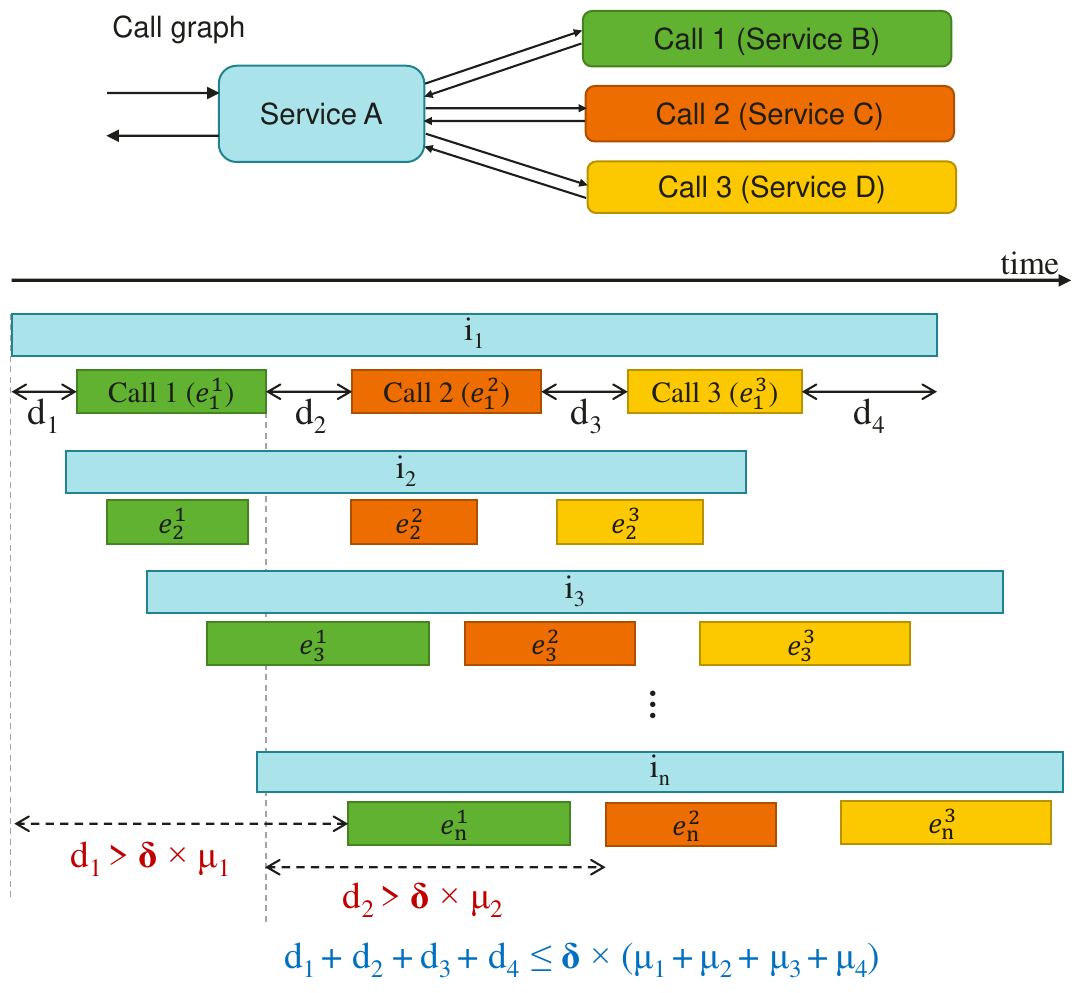}
	\caption{Call graph of microservice A and illustration of using adaptive thresholds based on average delay to filter egress spans of an ingress span in microservice A when identifying potential candidates.}
	\label{fig4}
\end{figure}

\textbf{Step 1 -- Identify Potential Candidates}: Finding candidates has a time complexity that grows exponentially with the number of concurrent requests and egress calls. This exponential growth significantly impacts the runtime needed to correlate spans in high-load systems. For example, in Fig. \ref{fig4}, microservice A receives an ingress request and communicates with three microservices: B, C, and D. In this case, a candidate for an ingress span in microservice A is a combination of three egress spans that occur during the ingress request. In high-load systems with hundreds or thousands of concurrent requests, simply filtering egress spans based on start and end times without a threshold can result in millions of potential combinations, making the process computationally expensive.

To address this, a threshold is required to prevent the inclusion of distant candidates that are unlikely to represent correct correlations. However, a fixed threshold is not optimal due to the variability in delays across service calls. Therefore, we propose an adaptive threshold based on a statistically informed heuristic—a simple yet effective approach to optimize both search time and candidate accuracy. In real-world systems, where delays often follow a positively skewed distribution (e.g., log-normal or exponential), the mean alone can serve as a coarse measure of central tendency, even when the variance is unknown. Thus, we define the threshold \(T\) as:
\begin{equation}
T_k = \delta\times\mu_k
\end{equation}
where \(\mu_k\) is the estimated mean of the \textit{k-th} delay and \(\delta\) is a configurable scaling factor. This multiplicative threshold retains candidates with delays smaller than \(T_k\), significantly reducing the number of candidates and the search time. For example, when searching for the first egress span to build potential candidates for the ingress span \(i_1\), only three closest egress spans \(e^1_1\), \(e^1_2\), and \(e^1_3\) are considered, as illustrated in Fig. \ref{fig4}. Empirical evaluation suggests that a setting of \(\delta=4\) captures the majority of plausible delays, even under long-tail distributions.

A practical challenge arises when computing \(\mu_k\) without knowing the exact correlations between ingress and egress spans. Interestingly, this can be addressed using a basic statistical property as the average of differences is equal to the difference of averages. That is, the average delay between two types of spans can be approximated as:
\begin{equation}
\mu_1 =  \frac{\sum_{i=1}^{N}t^s_i - \sum_{i=1}^{N}t^s_{e^i_1}}{N}
\end{equation}
where \(t^s_i\) and \(t^s_{e^1_i}\) re the start times of the ingress and first egress spans, respectively. This approach allows for estimation of \(\mu_1\), \(\mu_2\), etc., without prior knowledge of exact correlations. The total delay \(\mu_d\) can also be estimated using average durations, helping to filter out candidates with excessively large delays.

\begin{algorithm}
\caption{Cross-thread Span Correlation}
\label{alg:cap}
\begin{algorithmic}[1]
\Require{inSpans, egSpans, callGraph, $\delta$ = 4}
\Statex \textbf{Step 1: Identify potential candidates}
\For{$inSpan$ in $inSpans$}
    \State $topSpans \gets findspan(inSpan, egSpans, callGraph, \delta)$
\EndFor
\Statex \textbf{Step 2: Extract high-certainty set}
\State $high\_certainty \gets []$
\State $low\_certainty \gets []$
\State $diff\_threshold \gets 0.2$ \Comment{Difference Threshold is 20\%}
\For{$inSpan$ in $inSpans$}
    \State $diff = cal\_diff(topSpans)$
    \If{$diff \geq diff\_threshold$}
    \State $high\_certainty.append(inSpan)$
    \Else
    \State $low\_certainty.append(inSpan)$
    \EndIf
\EndFor
\Statex \textbf{Step 3: Build distribution models}
\State $dist\_models \gets buildModels(high\_certainty)$
\For{$span$ in $inSpans$}
    \State $calculatePDS(topCands[span], dist\_models)$
\EndFor
\Statex \textbf{Step 4: Greedy Assignment}
\State $sortedInSpans \gets sort(inSpans)$
\For{$span$ in $sortedSpans$}
    \If{$isConflicted(topCands[span])$}
    \State $resolveConflict(span)$
    \Else
    \State $correlateSpan(topCands[span])$
    \EndIf
\EndFor
\end{algorithmic}
\end{algorithm}

\textbf{Step 2 -- Extract High-certainty Set}: We observe that not all ingress spans are equally difficult to correlate, as requests with fewer candidates or distinct delays among candidates are generally easier to process. To take advantage of this observation, we compute a Central Deviation Score (CDS) for each candidate based on the average delays from the previous step, as follows:
\begin{equation}
C(i) = \sum_{k=1}^{n}\frac{\left| d_k - \mu_k \right|}{\mu_k}
\end{equation}
where \( d_k \) is \textit{k-th} delay between spans in the current candidate, and \( \mu_k \) is the corresponding average delay. A smaller \( C(i) \) value indicates closer alignment with the average delays. Ingress spans with candidates showing a significant difference (e.g., more than 20\%) between the smallest and second smallest CDS values, with no conflict from the top candidate, are classified as \textbf{high-certainty} spans, as the top candidate is likely to contain the correct egress spans (lines 9--10 in Algorithm \ref{alg:cap}) ). Conversely, spans that do not meet this criterion are classified as \textbf{low-certainty} spans. This classification rule is defined as:
\begin{equation}
\left\{
\begin{array}{ll}
    \frac{C(i_1) - C(i_0)}{C(i_0)} \geq D & \text{high-certainty ingress span} \\
    \frac{C(i_1) - C(i_0)}{C(i_0)} < D & \text{low-certainty ingress span}
    \end{array}
\right.
\end{equation}
where \( C(i_0) \), \( C(i_1) \), and \( D \) are the smallest and second smallest CDS values and the threshold to determine whether the difference is significant, respectively. The primary goal of this step is to extract these easier requests (i.e., the high-certainty set), which can then be leveraged for further analysis to refine the correlation process.

\textbf{Step 3 -- Build Distribution Models}: In this step, we need to find the distribution of the delays between spans based on high-certainty set. We begin by fitting several standard parametric distributions---specifically Gaussian (Normal), Log-Normal, and Exponential---to the delays extracted from the high-certainty set. To evaluate the adequacy of each fitted distribution, we employ a combination of goodness-of-fit tests, such as Kolmogorov-Smirnov \cite{massey1951kolmogorov}, Anderson-Darling \cite{stephens1974edf}, and Chi-Square \cite{snedecor1989statistical} tests, to quantitatively assess how well each parametric model aligns with the empirical data. These tests provide statistical measures that indicate the degree of conformity between the observed delays and the theoretical distributions. In addition to goodness-of-fit assessments, the Bayesian Information Criterion (BIC) \cite{bic} can serve as a model selection tool that balances the complexity of the model against its ability to fit the data.

If none of the standard parametric distributions demonstrate an adequate fit, as indicated by poor goodness-of-fit metrics and high BIC values, we employ a Gaussian Mixture Model (GMM) \cite{gmm}, as utilized in TraceWeaver \cite{traceweaver}. The GMM is a more flexible modeling approach that can capture complex data distributions by combining multiple Gaussian components. This allows for the accommodation of multi-modal patterns and greater variability in the delay data that single-parametric models may fail to represent effectively. To determine the optimal number of components \( C \), we evaluate GMMs with \( C \) ranging from 1 to 20 and select the model \( \text{GMM}(C) \) that achieves the lowest BIC score \cite{traceweaver}. Building a distribution model using the available dataset is a well-established process within the statistical domain, supported by robust methodologies and readily available implementations in various programming languages \cite{pedregosa2011scikit}.

Once the final distribution models are constructed, they are used to compute the total Probability Density Score (PDS) for each candidate based on their delays. For a given candidate \(C_i\), the total PDS \( P(C_i) \) is calculated as the sum of the log-probabilities of its individual delays under the fitted distributions:
\begin{equation}
P(C_i) = \sum_{k=1}^{n} \log f_k(d_k \mid \rho_k)
\end{equation}
where \(f_k\) represents the probability density function (PDF) of the chosen distribution, \(\rho_k\) denotes its parameters, and \(d_k\) is the observed \textit{k-th} delay. A higher PDS indicates a greater likelihood that the candidate represents the correct correlation between ingress and egress spans. This probabilistic ranking serves as the foundational factor for the subsequent process of assigning egress spans to ingress spans.

\textbf{Step 4 -- Greedy Assignment}: After scoring all candidates, ingress spans are sorted based on the difference between the highest and second-highest PDS among their candidates, aiming to prioritize ingress spans with the larger difference (i.e., spans with a high likelihood of correct correlation). By processing these spans first, the agent can progressively eliminate correlated spans from the candidate lists of the remaining ingress spans, thereby improving overall accuracy.

The agent assigns a candidate to each ingress span using a greedy approach in which candidate with the highest score will be selected. However, if this selection causes a conflict (i.e., an egress span is already included in another assigned combination), the agent performs conflict resolution for these conflicting ingress spans. Conflicts typically involve a small subset of ingress spans, making it feasible to search all possible combinations. The agent compares the overall scores of these combinations and selects the option that yields the highest total score for the conflicting spans. This additional conflict resolution step enhances the global optimization of the greedy assignment process and ultimately improves the accuracy of the correlation result.


\begin{figure}[!t]
	\centering
	\includegraphics[width=\linewidth]{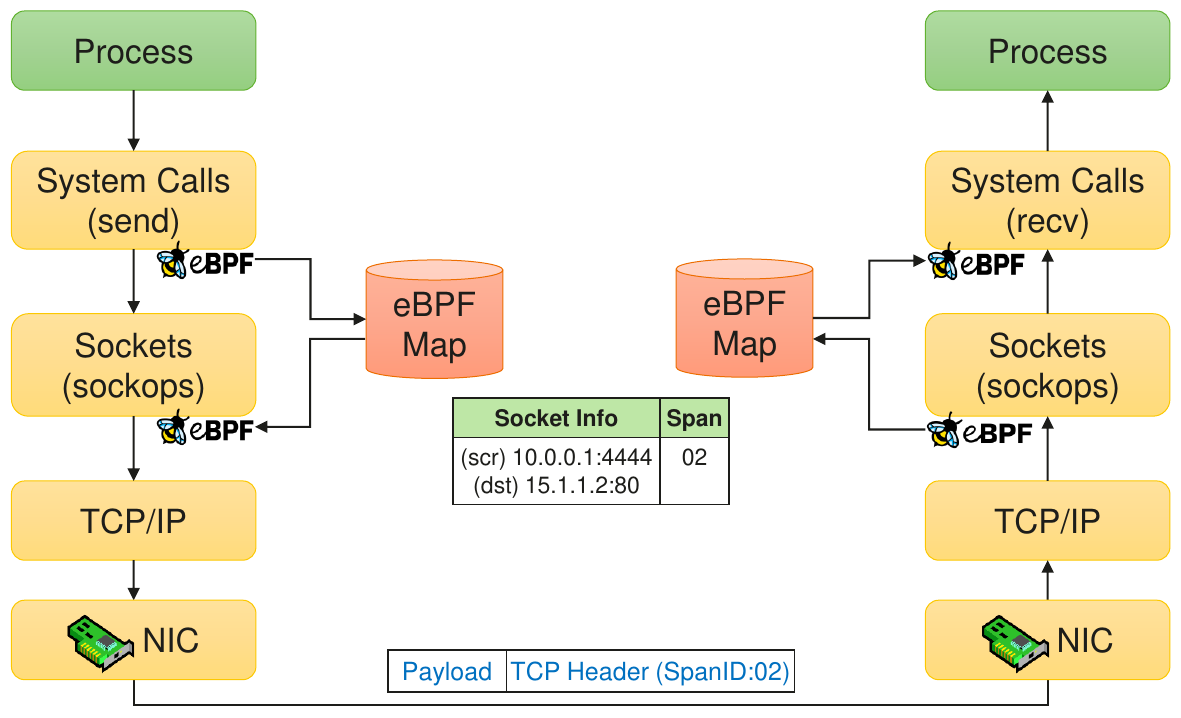}
	\caption{Cross-service span correlation through propagating span ID in the TCP header. The communication between two microservices is presented by two processes, from the sender to the receiver.}
	\label{fig5}
\end{figure}

\subsection{Cross-service Span Correlation}
To propagate metadata in the communication between two microservices without modifying the application layer, we must rely on a lower layer in the network stack. Since version 5.10, the Linux kernel has introduced helper functions (i.e., bpf\_reserve\_hdr\_opt, bpf\_store\_hdr\_opt, and bpf\_load\_hdr\_opt) in eBPF that allow reading and writing data into the option field of the TCP header. These functions were originally developed to support the implementation and testing of new network protocols and TCP congestion control algorithms. Leveraging these functions, we designed the cross-service span correlation feature in CrossTrace. Specifically, an eBPF program is attached to the socket layer of the operating system to propagate span IDs between microservices. An eBPF map is used to enable data sharing between eBPF programs in the system call layer (responsible for generating spans) and the socket layer (responsible for inserting and extracting span IDs), as illustrated in Fig. \ref{fig5}. 

At the start of a new egress span (e.g., row 3 in Table \ref{table:tb2}), the eBPF program in the system call layer writes the socket information (e.g., TCP tuple) and span ID to the shared map. When the request reaches the socket layer, the eBPF program retrieves the data from the shared map and embeds the span ID in the TCP header. On the receiving side, the eBPF program in the socket layer extracts the span ID from the TCP header and writes it, along with the socket information, back to the shared map. When the packet reaches the system call layer, the eBPF program retrieves the span ID from the shared map. This allows the newly created span at the receiving microservice to be correlated with its parent span in real time using the data embedded in the TCP header.

Notably, the eBPF helper functions used in our approach are enabled for each socket connection in the system by turning on a specific flag for that socket. This ensures that they are triggered exclusively for the communications of the microservices being traced rather than for all communications in the operating system. This mechanism helps minimize the overhead of using these functions on the running machine. The accuracy of this cross-service span correlation is comparable to that achieved by propagating trace context via request headers in instrumentation-based solutions, as both approaches rely on metadata propagation for span correlation.

\subsection{Trace Reconstruction}
Once spans are created and correlated, each eBPF agent submits them to the Trace Processor, a service responsible for reconstructing completed traces from correlated spans. Trace Processor is designed to enable compatibility between CrossTrace and third-party trace storage solutions. In many of these systems, a Trace ID is required in each span to construct complete traces for visualization. Trace reconstruction essentially involves identifying correlated spans and assigning them a unique trace ID. The Trace Processor can then forward these completed traces to storage and visualization systems. The Trace Processor can be implemented following OpenTelemetry specifications, enabling seamless integration with other open-source solutions. In large-scale systems, optimized search algorithms and caching techniques can be employed to accelerate the reconstruction process. However, this task is beyond the scope of this paper and is left for future work.

\section{Performance Evaluation}
\label{sec:4}
\subsection{Methodology}
\subsubsection{Testbed Setup} 
\begin{figure}[!t]
	\centering
	\includegraphics[width=0.9\linewidth]{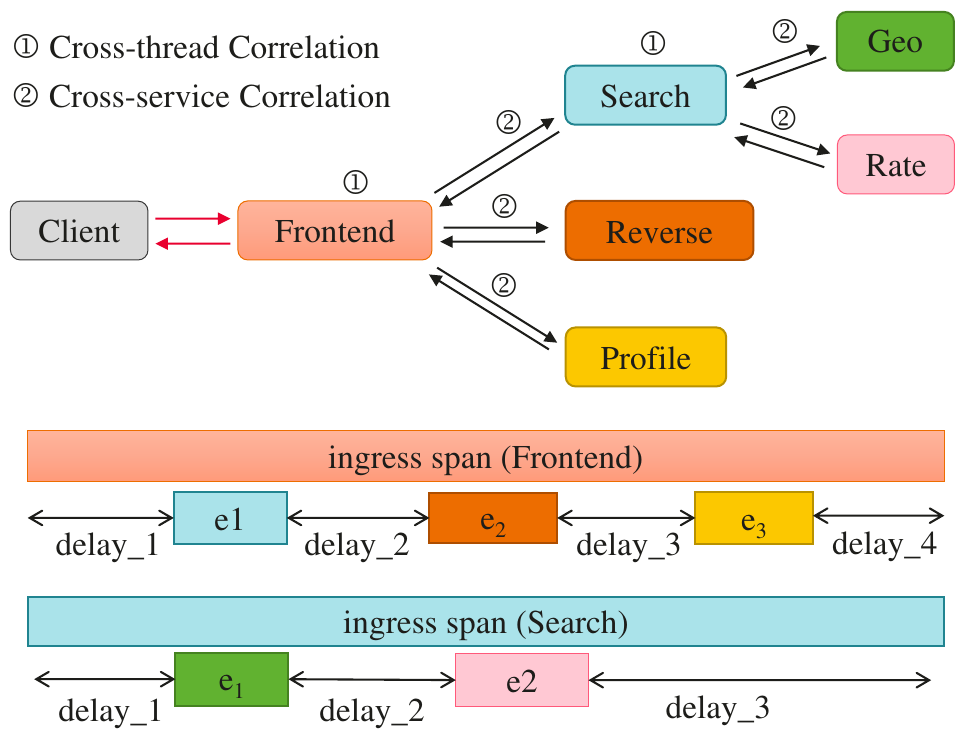}
	\caption{Service topology of the \textit{Hotel Reservation} application, the \textit{Frontend} and \textit{Search} are microservices involved in cross-thread span correlation.}
	\label{topo}
\end{figure}

The eBPF Agent is implemented in C for eBPF programs and in Go for user-space code. We evaluate the proposed solution using the \textit{Hotel Reservation} benchmark application from the DeathStarBench~\cite{deathstar} suite. This application is deployed via Docker Compose~\cite{docker} on a physical server equipped with an Intel Core i7 8700K processor, 64 GB of memory, and running Ubuntu 22.04 with Linux kernel version 6.5. We intentionally used an older machine to align with the requirement from the production team that only 5\% of the total CPU resources of the server are allocated to observability solutions. Figure \ref{topo} illustrates the service topology of the \textit{Hotel Reservation} application. We set \(\delta = 4\) for filtering egress spans when finding potential candidates and a threshold of \(D = 0.2\) (i.e., 20\% difference ratio) to extract the high-certainty set.

\subsubsection{Trace Dataset} Since CrossTrace relies solely on delays to correlate intra-service spans, its accuracy depends on the distinctness of spans. In other words, when two candidates have both similar start and end times, timing information alone becomes insufficient to distinguish between them. This scenario happens when request concurrency increases, resulting the higher likelihood of multiple requests occurring simultaneously. To evaluate accuracy under varying concurrency levels, we generated synthetic trace datasets from 10,000 independent requests. A Python program was used to adjust both the start and end times of spans within these requests, ensuring that their original durations and delays between spans were maintained. This adjustment allows the creation of a concurrent dataset, simulating high-load scenarios from the initially non-concurrent dataset. We evaluated six concurrency levels: 250, 500, 750, 1000, 1250, and 1500. These levels reflect realistic workloads for a single container in modern cloud systems, given that most cloud providers support up to 1000 concurrent requests per instance for container services~\cite{huawei,aws,gcp}.



\subsubsection{Comparison Objectives} The evaluation focuses on two primary aspects: the accuracy and efficiency of cross-thread span correlation and the overhead introduced by cross-service span correlation. We compare our solution against \textit{TraceWeaver}~\cite{traceweaver}, a recent approach that employs optimization techniques for span correlation. Other solutions, \textit{Deepflow} and \textit{Grafana Beyla}, are excluded from this evaluation, as they do not support cross-thread correlation, making them unsuitable benchmarks. For cross-service span correlation, our method directly propagates the span ID, ensuring inherent accuracy. Thus, the evaluation primarily focuses on the overhead introduced by the eBPF program attached to the socket layer for span ID propagation.

\subsection{Accuracy of Cross-Thread Span Correlation}
\begin{figure}[!t]
	\centering
	\includegraphics[width=\linewidth]{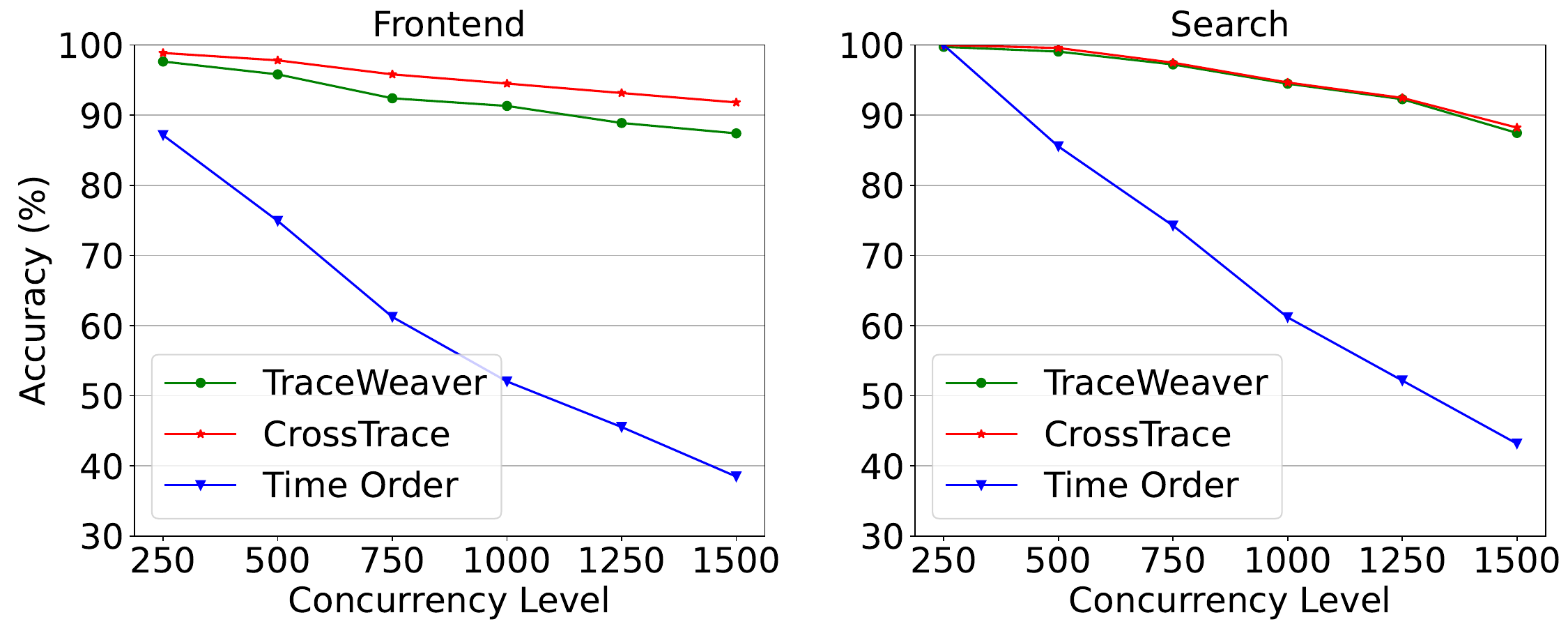}
	\caption{Accuracy of CrossTrace and TraceWeaver in \mbox{correlating} spans of \textit{Frontend} and \textit{Search} microservices under different concurrency levels.}
	\label{fig7}
\end{figure}

At low concurrency levels (e.g., 250–500 concurrent requests), CrossTrace achieves a correlation accuracy of 98\%, comparable to that of TraceWeaver. As the concurrency level increases to over 1000 concurrent requests, CrossTrace continues to perform well, maintaining an accuracy of over 90\% for the \textit{Frontend} microservice (slightly higher than TraceWeaver) and 88\% for the \textit{Search} microservice (similar to TraceWeaver), as shown in Figure \ref{fig7}. In contrast, a simpler approach that selects the closest spans suffers a significant decline in accuracy as concurrency increases. We observe that the accuracy of the \textit{Search} microservice is slightly lower than that of the \textit{Frontend} microservice due to the magnitude of the last delay (i.e., \textit{delay\_2}), which is over 10000~$\mu$s (10 ms). A larger range of delays results in a greater number of potential candidates, increasing the likelihood of incorrect egress spans with higher PDS values being selected, which leads to incorrect correlations.

\begin{figure}[!t]
	\centering
	\includegraphics[width=\linewidth]{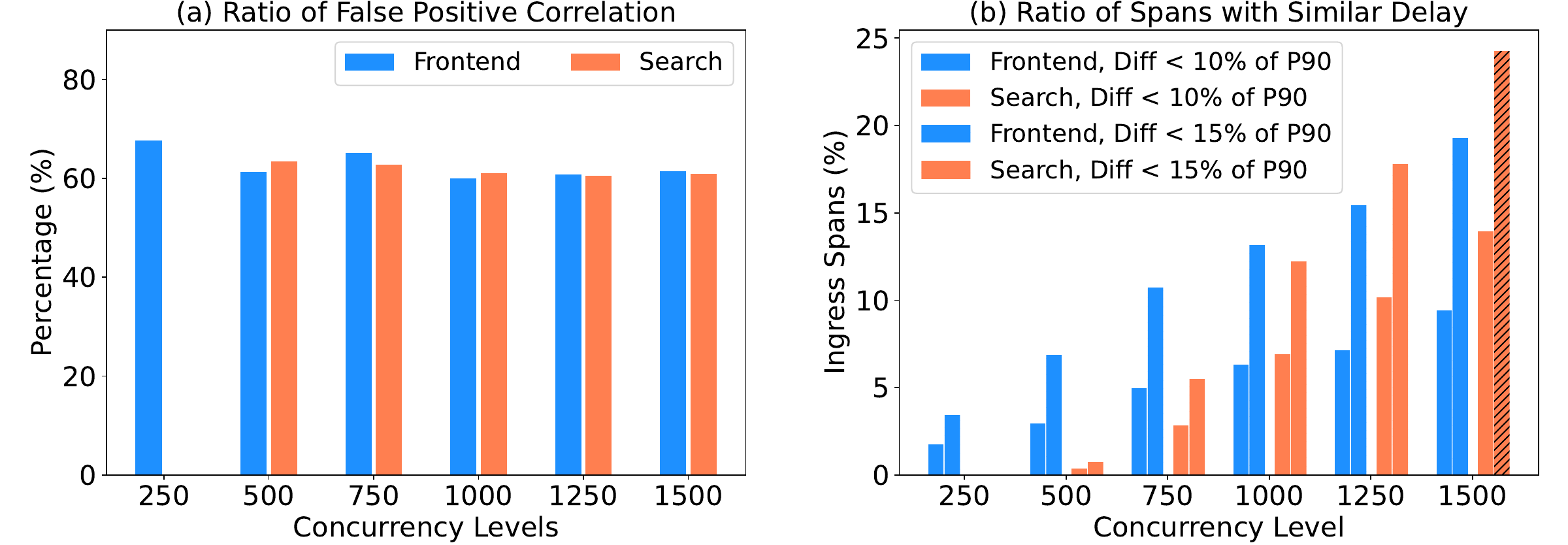}
	\caption{(a) Percentage of incorrect correlations where the correct candidate has a smaller Probability Density Score than the top candidate. (b) Ratio of ingress spans where the differences in delays of their egress spans are less than 10\% and 15\% of the P90 delays.}
	\label{fig8}
\end{figure}

Further investigation into incorrect correlation cases reveals that about 60\% of these wrong correlations involve ingress spans where the correct candidate has a smaller PDS than that of the top candidate, as shown in Figure \ref{fig8}a. This observation indicates that the correlation algorithm performs as designed, prioritizing candidates with the highest PDS. However, it also highlights that the top candidate is not always the correct one. For instance, in requests where delays fall into the tail of the distribution, the correct correlation may have a smaller PDS. Additionally, Figure \ref{fig8}b spresents the proportion of ingress spans for which the differences in delays between candidate egress spans are less than 10\% and 15\% of the P90 delay. This data reflects the distinctness level of spans in our evaluation. When span timestamps are extremely close, the delay differences between candidates become minimal, resulting in multiple candidates having similarly high PDS values. These findings suggests that the accuracy of CrossTrace is approaching its upper bound, as most remaining incorrect correlations arise from inherently ambiguous cases, leaving limited room for further improvement.

In summary, the accuracy of cross-thread span correlation depends heavily on the distinctiveness of delays between spans. When the delays of candidates are highly similar, the accuracy of CrossTrace may decrease. However, such scenarios may happen in extremely high-load systems, allowing CrossTrace to remain effective for other typical applications. We discuss a potential solution to address these ambiguous correlations in Section \ref{sec:5}.

\subsection{Runtime of Cross-Thread Span Correlation}
\begin{figure}[!t]
	\centering
	\includegraphics[width=\linewidth]{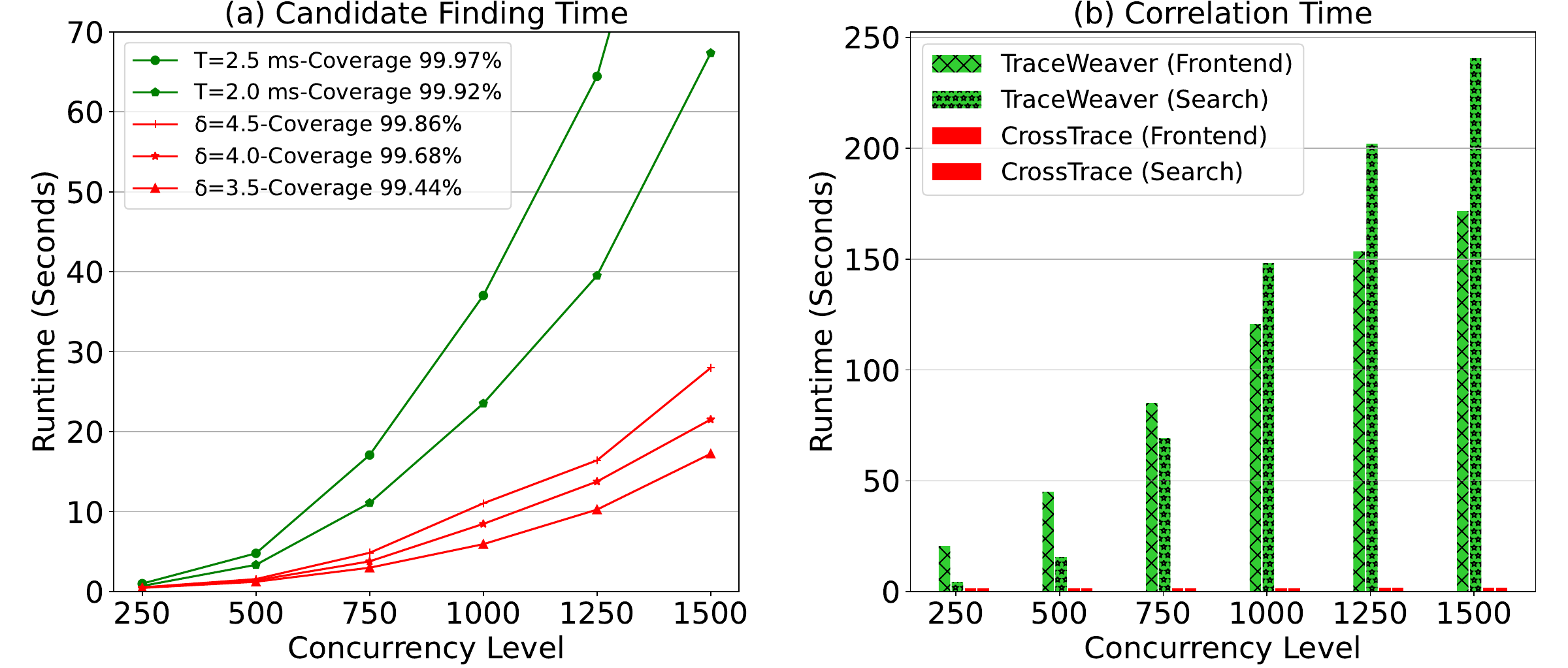}
	\caption{(a) Time to find candidates using different fixed thresholds (i.e., T=2.0 ms and T=2.5 ms) and adaptive thresholds based on average delays. (b) Time to determine final correlations for ingress and egress spans in TraceWeaver and CrossTrace.}
	\label{fig9}
\end{figure}

To evaluate the runtime performance of CrossTrace compared to TraceWeaver, an offline implementation of CrossTrace was developed in Python---the same language used for implementing TraceWeaver---to ensure fair comparisons. Runtime is primarily divided into two parts: candidate finding time and correlation time. The time required to find potential candidates increases with higher concurrency levels due to the larger search space. Although the original design of TraceWeaver did not incorporate a threshold mechanism, the candidate-finding step can be implemented in the same way for both solutions. Therefore, instead of directly comparing TraceWeaver and CrossTrace for this step, we focus on evaluating the impact of different threshold mechanisms (i.e., fixed threshold and adaptive threshold) on runtime. Among these mechanisms, the adaptive threshold approach---based on a multiple \(\delta\) of the average delay---used in CrossTrace proves to be the most optimal. The fixed threshold mechanism, on the other hand, results in a significantly higher runtime, which is 2--4$\times$ longer than the adaptive threshold, as shown in Fig. \ref{fig9}a. For high-tailed delay applications, increasing the value of \(\delta\) slightly increases runtime while achieving higher coverage. 

Fig.\ref{fig9}b compares the correlation time between CrossTrace and TraceWeaver. CrossTrace maintains a consistent correlation time of less than 2~seconds, regardless of concurrency levels, due to its efficient greedy assignment approach. In contrast, TraceWeaver's runtime increases significantly as concurrency levels increase, driven by the computational overhead of its optimization process. Overall, TraceWeaver's correlation time is 20--200$\times$ longer than CrossTrace, depending on the concurrency level. In summary, these results highlight the efficiency of using adaptive thresholds and the greedy correlation process in CrossTrace, making it suitable for high-load systems.

\begin{figure}[!t]
	\centering
	\includegraphics[width=0.85\linewidth]{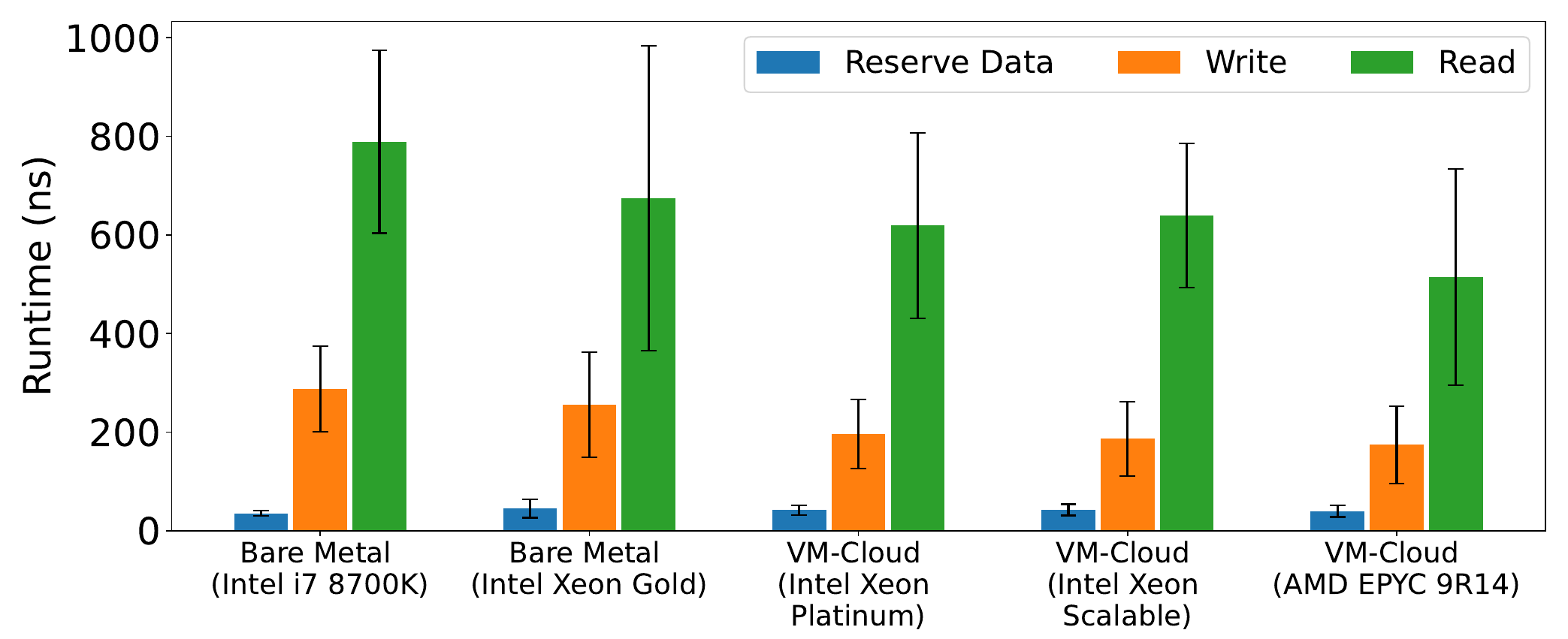}
	\caption{The runtime overhead of eBPF helper functions at the socket layer for propagating span ID.}
	\label{fig10}
\end{figure}

\subsection{Overhead in Cross-Service Span Correlation}
The runtime overhead of eBPF programs attached to system calls has been studied in prior research \cite{deepflow, ebpftime, ebpfmap}, with overall results showing that overhead of capturing spans is small. Here, we specifically evaluate the runtime overhead of eBPF helper functions at the socket layer used to propagate span IDs. These functions involve reserving space and performing read/write operations between eBPF maps and the TCP header using eBPF helper functions. The overhead of these operations is measured by tracking the execution time across 50,000 requests. These tests were conducted on several common CPUs running on bare metal and virtual machines in the cloud, as shown in Fig.~\ref{fig10}.

The results show that reserving space in the TCP header typically takes less than 50~ns, writing data to the TCP header takes less than 300~ns, and reading takes more than twice as long as writing, at over 600~ns. On systems with more powerful CPUs, these times can be reduced, demonstrating the potential for even lower overhead in modern cloud infrastructures. It is important to note that these eBPF helper functions are executed only for requests originating from traced microservices, not for all communications on the machine. Overall, these results indicate that the overhead introduced by the eBPF program for cross-service span correlation is minimal, and this approach can be safely applied to high-throughput workloads.

\section{Limitations and Discussion}
\label{sec:5}
The main limitation of cross-thread span correlation lies in its reliance on delay information to infer relationships between spans. When the differences among delays become extremely small, it becomes uncertain which correlation is correct. To improve the accuracy and usefulness of the resulting traces, the greedy algorithm can be extended to return multiple top candidates, increasing the likelihood of including the correct egress span. 

By default, each egress span is associated with only one ingress span (i.e., a one-to-one mapping). As such, CrossTrace currently selects a single candidate as the final correlation, even when multiple plausible candidates are identified. To return more than one correlation result while still respecting the one-to-one mapping constraint, the algorithm the algorithm creates duplicates of egress spans to support additional correlations. For instance, consider two ingress spans, \(i_1\) and \(i_2\), with their corresponding egress spans being \(e_1\) and \(e_2\). If \(e_1\) and \(e_2\) occur at nearly the same time, they may have similar delays and PDS values, making them the suitable candidates for both \(i_1\) and \(i_2\). Instead of selecting only one egress span for each ingress span, CrossTrace duplicates the egress spans (e.g., \(e_{1a}\) and \(e_{2a}\)), and returns multiple correlation results for each ingress span (i.e., \(i_1 \rightarrow e_1\), \(i_1 \rightarrow e_2\), \(i_2 \rightarrow e_{1a}\), \(i_2 \rightarrow e_{2a}\)).

\begin{figure}[!t]
	\centering
	\includegraphics[width=\linewidth]{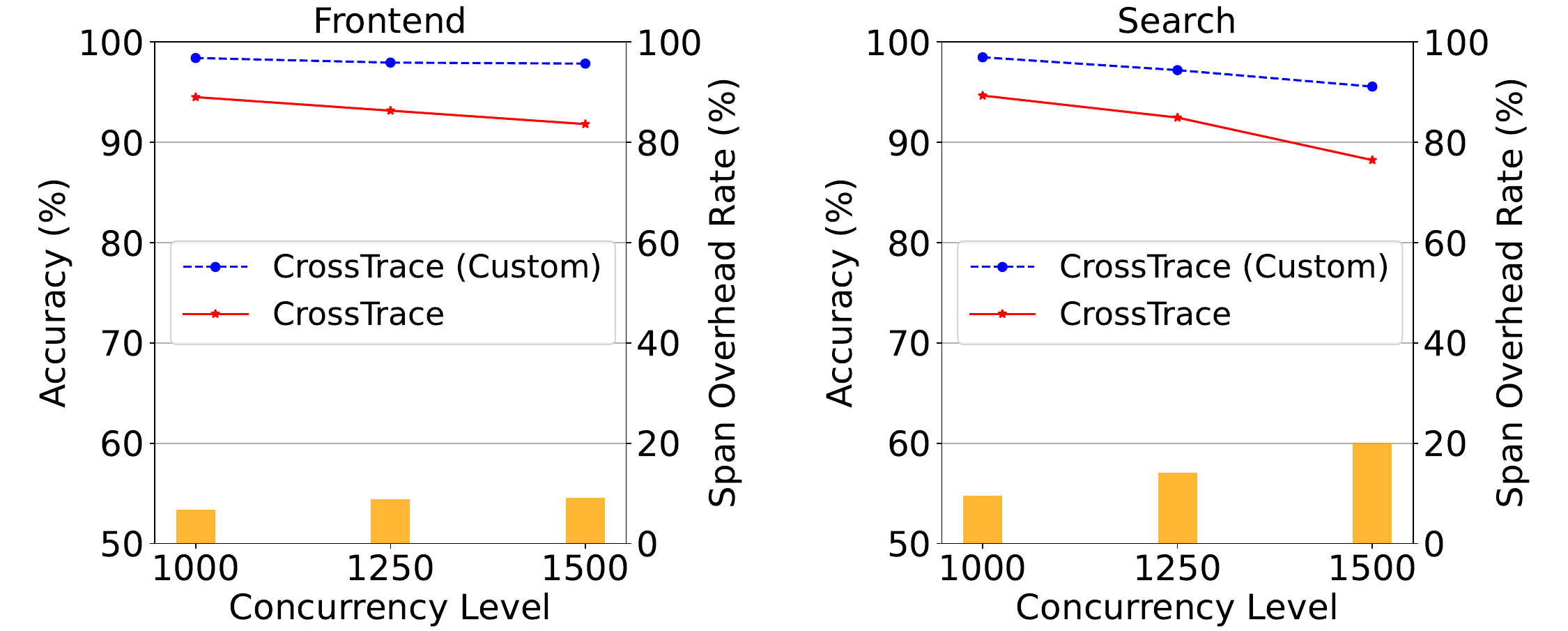}
	\caption{Accuracy and span overhead rate in a customized version of CrossTrace that returns multiple correlations for top candidates.}
	\label{fig11}
\end{figure}

While duplicating spans introduces additional overhead in the storage system, it enhances troubleshooting by ensuring that the final traces contain all necessary spans for further diagnosis and analysis. As shown in Figure~\ref{fig11}, returning multiple correlation results enables CrossTrace to achieve over 95\% accuracy, with approximately 10\% and 20\% increases in the number of spans for the \textit{Frontend} and \textit{Search} microservices, respectively. The flexibility of the greedy algorithm allows it to be tailored for specific use cases. For example, returning multiple correlation results can be selectively enabled for abnormal requests (e.g., those with HTTP 502 responses or unusually high latency). This ensures that critical traces contain comprehensive information for root cause analysis, while minimizing storage overhead for normal traffic.

\section{Conclusion}
\label{sec:6}
This paper presents CrossTrace, an eBPF-based distributed tracing solution that supports zero-code instrumentation while addressing critical challenges in span correlation both within and across services. For intra-service span correlation, CrossTrace introduces a greedy algorithm that infers causal relationships based on delay patterns, eliminating dependence on thread identifiers. For inter-service correlation, it embeds span ID into TCP headers using eBPF, avoiding modifications to application-layer protocols. CrossTrace is designed with real-world production constraints in mind, offering a practical balance between accuracy and performance under high-concurrency conditions. Although relying solely on delay information may limit the ability to disambiguate spans with near-identical timestamps, our evaluation shows that CrossTrace consistently achieves high correlation accuracy across realistic workload scenarios. These results highlight its effectiveness in enhancing observability and debugging capabilities in microservice applications. Future work will focus on evaluating CrossTrace in production deployments to further assess its impact on microservice troubleshooting.




\bibliographystyle{IEEEtran}

\begin{thebibliography}{10}
\providecommand{\url}[1]{#1}
\csname url@samestyle\endcsname
\providecommand{\newblock}{\relax}
\providecommand{\bibinfo}[2]{#2}
\providecommand{\BIBentrySTDinterwordspacing}{\spaceskip=0pt\relax}
\providecommand{\BIBentryALTinterwordstretchfactor}{4}
\providecommand{\BIBentryALTinterwordspacing}{\spaceskip=\fontdimen2\font plus
\BIBentryALTinterwordstretchfactor\fontdimen3\font minus \fontdimen4\font\relax}
\providecommand{\BIBforeignlanguage}[2]{{%
\expandafter\ifx\csname l@#1\endcsname\relax
\typeout{** WARNING: IEEEtran.bst: No hyphenation pattern has been}%
\typeout{** loaded for the language `#1'. Using the pattern for}%
\typeout{** the default language instead.}%
\else
\language=\csname l@#1\endcsname
\fi
#2}}
\providecommand{\BIBdecl}{\relax}
\BIBdecl

\bibitem{9678534}
T.~Yang, J.~Shen, Y.~Su, X.~Ling, Y.~Yang, and M.~R. Lyu, ``Aid: Efficient prediction of aggregated intensity of dependency in large-scale cloud systems,'' in \emph{2021 36th IEEE/ACM International Conference on Automated Software Engineering (ASE)}, 2021, pp. 653--665.

\bibitem{mystery}
M.~Chow, D.~Meisner, J.~Flinn, D.~Peek, and T.~F. Wenisch, ``The mystery machine: end-to-end performance analysis of large-scale internet services,'' in \emph{Proceedings of the 11th USENIX Conference on Operating Systems Design and Implementation}, ser. OSDI'14.\hskip 1em plus 0.5em minus 0.4em\relax USA: USENIX Association, 2014, p. 217–231.

\bibitem{trace-abnormally}
M.~Panahandeh, A.~Hamou-Lhadj, M.~Hamdaqa, and J.~Miller, ``Serviceanomaly: An anomaly detection approach in microservices using distributed traces and profiling metrics,'' \emph{Journal of Systems and Software}, vol. 209, p. 111917, 2024.

\bibitem{tracing-debug}
Z.~{Purfallah Mazraemolla} and A.~Rasoolzadegan, ``An effective failure detection method for microservice-based systems using distributed tracing data,'' \emph{Engineering Applications of Artificial Intelligence}, vol. 133, p. 108558, 2024.

\bibitem{tracing-debug2}
C.~Zhang, Z.~Dong, X.~Peng, B.~Zhang, and M.~Chen, ``Trace-based multi-dimensional root cause localization of performance issues in microservice systems,'' in \emph{Proceedings of the IEEE/ACM 46th International Conference on Software Engineering}, ser. ICSE '24.\hskip 1em plus 0.5em minus 0.4em\relax New York, NY, USA: Association for Computing Machinery, 2024.

\bibitem{otel}
\BIBentryALTinterwordspacing
OpenTelemetry, ``High-quality, ubiquitous, and portable telemetry to enable effective observability.'' 2024. [Online]. Available: \url{https://opentelemetry.io}
\BIBentrySTDinterwordspacing

\bibitem{ebpf-http}
C.~Liu, Z.~Cai, B.~Wang, Z.~Tang, and J.~Liu, ``A protocol-independent container network observability analysis system based on ebpf,'' in \emph{2020 IEEE 26th International Conference on Parallel and Distributed Systems (ICPADS)}.\hskip 1em plus 0.5em minus 0.4em\relax IEEE, 2020, pp. 697--702.

\bibitem{bpf}
S.~McCanne and V.~Jacobson, ``The bsd packet filter: a new architecture for user-level packet capture,'' in \emph{Proceedings of the USENIX Winter 1993 Conference Proceedings on USENIX Winter 1993 Conference Proceedings}, ser. USENIX'93.\hskip 1em plus 0.5em minus 0.4em\relax USA: USENIX Association, 1993, p.~2.

\bibitem{ebpf-insight}
L.~Song and J.~Li, ``ebpf: Pioneering kernel programmability and system observability - past, present, and future insights,'' in \emph{2024 3rd International Conference on Artificial Intelligence and Computer Information Technology (AICIT)}.\hskip 1em plus 0.5em minus 0.4em\relax IEEE, 2024, pp. 1--10.

\bibitem{ebpf-rise}
C.~Cassagnes, L.~Trestioreanu, C.~Joly, and R.~State, ``The rise of ebpf for non-intrusive performance monitoring,'' in \emph{NOMS 2020 - 2020 IEEE/IFIP Network Operations and Management Symposium}, 2020, pp. 1--7.

\bibitem{deepflow}
J.~Shen, H.~Zhang, Y.~Xiang, X.~Shi, X.~Li, Y.~Shen, Z.~Zhang, Y.~Wu, X.~Yin, J.~Wang, M.~Xu, Y.~Li, J.~Yin, J.~Song, Z.~Li, and R.~Nie, ``Network-centric distributed tracing with deepflow: Troubleshooting your microservices in zero code,'' in \emph{Proceedings of the ACM SIGCOMM 2023 Conference}, ser. ACM SIGCOMM '23.\hskip 1em plus 0.5em minus 0.4em\relax New York, NY, USA: Association for Computing Machinery, 2023, p. 420–437.

\bibitem{grafanabeyla}
\BIBentryALTinterwordspacing
G.~Beyla, ``Open source zero-code automatic instrumentation with ebpf and opentelemetry.'' 2024. [Online]. Available: \url{https://github.com/grafana/beyla}
\BIBentrySTDinterwordspacing

\bibitem{beyla-limitation}
\BIBentryALTinterwordspacing
Grafana, ``Distributed traces with grafana beyla,'' 2024. [Online]. Available: \url{https://grafana.com/docs/beyla/latest/distributed-traces}
\BIBentrySTDinterwordspacing

\bibitem{ebpf-threat}
\BIBentryALTinterwordspacing
J.~Kelly, J.~Callaghan, and A.~Martin, ``ebpf security threat model,'' eBPF Foundation, Tech. Rep., 2024. [Online]. Available: \url{https://www.linuxfoundation.org/hubfs/eBPF/ControlPlane — eBPF Security Threat Model.pdf}
\BIBentrySTDinterwordspacing

\bibitem{goroutine}
\BIBentryALTinterwordspacing
Golang, ``Goroutines in golang,'' 2024. [Online]. Available: \url{https://golangdocs.com/goroutines-in-golang}
\BIBentrySTDinterwordspacing

\bibitem{virtualthread}
\BIBentryALTinterwordspacing
Oracle, ``Virtual threads,'' 2024. [Online]. Available: \url{https://docs.oracle.com/en/java/javase/21/core/virtual-threads.html}
\BIBentrySTDinterwordspacing

\bibitem{traceweaver}
S.~Ashok, V.~Harsh, B.~Godfrey, R.~Mittal, S.~Parthasarathy, and L.~Shwartz, ``Traceweaver: Distributed request tracing for microservices without application modification,'' in \emph{Proceedings of the ACM SIGCOMM 2024 Conference}, ser. ACM SIGCOMM '24.\hskip 1em plus 0.5em minus 0.4em\relax New York, NY, USA: Association for Computing Machinery, 2024, p. 828–842.

\bibitem{istio}
\BIBentryALTinterwordspacing
Istio, ``The istio service mesh,'' 2024. [Online]. Available: \url{https://istio.io}
\BIBentrySTDinterwordspacing

\bibitem{Linkerd}
\BIBentryALTinterwordspacing
Linkerd, ``Service mesh,'' 2024. [Online]. Available: \url{https://linkerd.io}
\BIBentrySTDinterwordspacing

\bibitem{sambasivan2014so}
\BIBentryALTinterwordspacing
R.~R. Sambasivan, R.~Fonseca, I.~Shafer, and G.~R. Ganger, ``So, you want to trace your distributed system? key design insights from years of practical experience,'' Parallel Data Lab., Carnegie Mellon Univ., Pittsburgh, PA, USA, Tech. Rep., 2014. [Online]. Available: \url{https://www.pdl.cmu.edu/ftp/SelfStar/CMU-PDL-14-102.pdf}
\BIBentrySTDinterwordspacing

\bibitem{dapper}
\BIBentryALTinterwordspacing
B.~H. Sigelman, L.~A. Barroso, M.~Burrows, P.~Stephenson, M.~Plakal, D.~Beaver, S.~Jaspan, and C.~Shanbhag, ``Dapper, a large-scale distributed systems tracing infrastructure,'' Google, Inc., Tech. Rep., 2010. [Online]. Available: \url{http://research.google.com/archive/papers/dapper-2010-1.pdf}
\BIBentrySTDinterwordspacing

\bibitem{zipkin}
\BIBentryALTinterwordspacing
Zipkin, ``A distributed tracing system.'' 2024. [Online]. Available: \url{https://zipkin.io}
\BIBentrySTDinterwordspacing

\bibitem{jaeger}
\BIBentryALTinterwordspacing
Jaeger, ``Jaeger: open source, distributed tracing platform.'' 2024. [Online]. Available: \url{https://www.jaegertracing.io}
\BIBentrySTDinterwordspacing

\bibitem{opentracing}
\BIBentryALTinterwordspacing
OpenTracing, ``The opentracing project.'' 2024. [Online]. Available: \url{https://opentracing.io}
\BIBentrySTDinterwordspacing

\bibitem{opencensus}
\BIBentryALTinterwordspacing
OpenCensus, ``The opencensus project.'' 2024. [Online]. Available: \url{https://opencensus.io}
\BIBentrySTDinterwordspacing

\bibitem{sidecartiwari2017}
\BIBentryALTinterwordspacing
A.~Tiwari, ``A sidecar for your service mesh,'' \emph{Abhishek Tiwari}, 2017. [Online]. Available: \url{https://www.abhishek-tiwari.com/pdf/a-sidecar-for-your-service-mesh.pdf}
\BIBentrySTDinterwordspacing

\bibitem{Kubernetes}
\BIBentryALTinterwordspacing
Kubernetes, ``Production-grade container orchestration,'' 2024. [Online]. Available: \url{https://kubernetes.io}
\BIBentrySTDinterwordspacing

\bibitem{sidecar-drawback}
X.~Zhu, G.~She, B.~Xue, Y.~Zhang, Y.~Zhang, X.~K. Zou, X.~Duan, P.~He, A.~Krishnamurthy, M.~Lentz, D.~Zhuo, and R.~Mahajan, ``Dissecting overheads of service mesh sidecars,'' in \emph{Proceedings of the 2023 ACM Symposium on Cloud Computing}, ser. SoCC '23.\hskip 1em plus 0.5em minus 0.4em\relax New York, NY, USA: Association for Computing Machinery, 2023, p. 142–157.

\bibitem{ebpf-sidecar}
D.~Soldani, P.~Nahi, H.~Bour, S.~Jafarizadeh, M.~F. Soliman, L.~Di~Giovanna, F.~Monaco, G.~Ognibene, and F.~Risso, ``ebpf: A new approach to cloud-native observability, networking and security for current (5g) and future mobile networks (6g and beyond),'' \emph{IEEE Access}, vol.~11, pp. 57\,174--57\,202, 2023.

\bibitem{xtrace}
R.~Fonseca, G.~Porter, R.~H. Katz, S.~Shenker, and I.~Stoica, ``X-trace: a pervasive network tracing framework,'' in \emph{Proceedings of the 4th USENIX Conference on Networked Systems Design \& Implementation}, ser. NSDI'07.\hskip 1em plus 0.5em minus 0.4em\relax USA: USENIX Association, 2007, p.~20.

\bibitem{w3c}
\BIBentryALTinterwordspacing
W.~W. W.~C. (W3C), ``Trace context,'' 2024. [Online]. Available: \url{https://www.w3.org/TR/trace-context}
\BIBentrySTDinterwordspacing

\bibitem{envoy}
\BIBentryALTinterwordspacing
Envoy, ``Envoy proxy,'' 2024. [Online]. Available: \url{https://www.envoyproxy.io}
\BIBentrySTDinterwordspacing

\bibitem{vpath}
B.~C. Tak, C.~Tang, C.~Zhang, S.~Govindan, B.~Urgaonkar, and R.~N. Chang, ``vpath: precise discovery of request processing paths from black-box observations of thread and network activities,'' in \emph{Proceedings of the 2009 Conference on USENIX Annual Technical Conference}, ser. USENIX'09.\hskip 1em plus 0.5em minus 0.4em\relax USA: USENIX Association, 2009, p.~19.

\bibitem{wap5}
P.~Reynolds, J.~L. Wiener, J.~C. Mogul, M.~K. Aguilera, and A.~Vahdat, ``Wap5: black-box performance debugging for wide-area systems,'' in \emph{Proceedings of the 15th International Conference on World Wide Web}, ser. WWW '06.\hskip 1em plus 0.5em minus 0.4em\relax New York, NY, USA: Association for Computing Machinery, 2006, p. 347–356.

\bibitem{timing-infer}
T.~Wang, C.-s. Perng, T.~Tao, C.~Tang, E.~So, C.~Zhang, R.~Chang, and L.~Liu, ``A temporal data-mining approach for discovering end-to-end transaction flows,'' in \emph{2008 IEEE International Conference on Web Services}, 2008, pp. 37--44.

\bibitem{ebpf-net}
P.~G. Kannan, S.~M. Gupta, D.~Behl, E.~Raichstein, and J.~Takvorian, ``Designing a lightweight network observability agent for cloud applications,'' in \emph{Passive and Active Measurement}, P.~Richter, V.~Bajpai, and E.~Carisimo, Eds.\hskip 1em plus 0.5em minus 0.4em\relax Cham: Springer Nature Switzerland, 2024, pp. 262--276.

\bibitem{ebpf-ViperProbe}
J.~Levin and T.~A. Benson, ``Viperprobe: Rethinking microservice observability with ebpf,'' in \emph{2020 IEEE 9th International Conference on Cloud Networking (CloudNet)}, 2020, pp. 1--8.

\bibitem{callgraph}
S.~Luo, H.~Xu, C.~Lu, K.~Ye, G.~Xu, L.~Zhang, J.~He, and C.~Xu, ``An in-depth study of microservice call graph and runtime performance,'' \emph{IEEE Transactions on Parallel and Distributed Systems}, vol.~33, no.~12, pp. 3901--3914, 2022.

\bibitem{massey1951kolmogorov}
F.~J. Massey~Jr, ``The kolmogorov-smirnov test for goodness of fit,'' \emph{Journal of the American statistical Association}, vol.~46, no. 253, pp. 68--78, 1951.

\bibitem{stephens1974edf}
M.~A. Stephens, ``Edf statistics for goodness of fit and some comparisons,'' \emph{Journal of the American statistical Association}, vol.~69, no. 347, pp. 730--737, 1974.

\bibitem{snedecor1989statistical}
G.~W. Snedecor and W.~G. Cochran, ``Statistical methods, eight edition,'' \emph{Iowa state University press, Ames, Iowa}, vol. 1191, no.~2, 1989.

\bibitem{bic}
A.~A. Neath and J.~E. Cavanaugh, ``The bayesian information criterion: background, derivation, and applications,'' \emph{Wiley Interdisciplinary Reviews: Computational Statistics}, vol.~4, no.~2, pp. 199--203, 2012.

\bibitem{gmm}
D.~A. Reynolds \emph{et~al.}, ``Gaussian mixture models.'' \emph{Encyclopedia of biometrics}, vol. 741, no. 659-663, 2009.

\bibitem{pedregosa2011scikit}
F.~Pedregosa, G.~Varoquaux, A.~Gramfort, V.~Michel, B.~Thirion, O.~Grisel, M.~Blondel, P.~Prettenhofer, R.~Weiss, V.~Dubourg \emph{et~al.}, ``Scikit-learn: Machine learning in python,'' \emph{the Journal of machine Learning research}, vol.~12, pp. 2825--2830, 2011.

\bibitem{deathstar}
Y.~G. et~al., ``An open-source benchmark suite for microservices and their hardware-software implications for cloud \& edge systems,'' in \emph{Proceedings of the Twenty-Fourth International Conference on Architectural Support for Programming Languages and Operating Systems}, ser. ASPLOS '19.\hskip 1em plus 0.5em minus 0.4em\relax New York, NY, USA: Association for Computing Machinery, 2019, p. 3–18.

\bibitem{docker}
\BIBentryALTinterwordspacing
D.~Compose, ``Docker compose,'' 2024. [Online]. Available: \url{https://docs.docker.com/compose}
\BIBentrySTDinterwordspacing

\bibitem{huawei}
\BIBentryALTinterwordspacing
Huawei, ``Huawei cloud functiongraph.'' 2024. [Online]. Available: \url{https://support.huaweicloud.com/intl/en-us/usermanual-functiongraph/functiongraph\_01\_0303.html}
\BIBentrySTDinterwordspacing

\bibitem{aws}
\BIBentryALTinterwordspacing
Amazon, ``Amazon web services lambda,'' 2024. [Online]. Available: \url{https://docs.aws.amazon.com/lambda/latest/dg/lambda-concurrency.html}
\BIBentrySTDinterwordspacing

\bibitem{gcp}
\BIBentryALTinterwordspacing
Google, ``Google cloud run,'' 2024. [Online]. Available: \url{https://cloud.google.com/run/docs/about-concurrency}
\BIBentrySTDinterwordspacing

\bibitem{ebpftime}
R.~Sahu and D.~Williams, ``Enabling bpf runtime policies for better bpf management,'' in \emph{Proceedings of the 1st Workshop on EBPF and Kernel Extensions}, ser. eBPF '23.\hskip 1em plus 0.5em minus 0.4em\relax New York, NY, USA: Association for Computing Machinery, 2023, p. 49–55.

\bibitem{ebpfmap}
C.~Liu, B.~Tak, and L.~Wang, ``Understanding performance of ebpf maps,'' in \emph{Proceedings of the ACM SIGCOMM 2024 Workshop on EBPF and Kernel Extensions}, ser. eBPF '24.\hskip 1em plus 0.5em minus 0.4em\relax New York, NY, USA: Association for Computing Machinery, 2024, p. 9–15.

\end{thebibliography}

\end{document}